\documentclass[12pt]{article}
\usepackage{amsfonts}
\usepackage{amssymb}
\usepackage{amsmath}
\usepackage{bbm}
\usepackage{graphicx}
\usepackage{cite}
\usepackage{color}

\begin{document}
\begin{flushleft}
{\Large
\textbf{Data reliability in complex directed networks}
}
\\
Joaqu\'{\i}n Sanz$^{1,2\ast}$,
Emanuele Cozzo$^{1,2}$
Yamir Moreno$^{1,2}$
\\
\bf{1} Institute for Biocomputation and Physics of Complex Systems, University of Zaragoza, Zaragoza 50018, Spain.
\\
\bf{2} Department of Theoretical Physics, Faculty of Sciences, University of Zaragoza, Zaragoza 50009, Spain.
\\
$\ast$ E-mail: jsanz@bifi.es
\end{flushleft}

\begin{abstract}

The availability of data from many different sources and fields of science has made it possible to map out an increasing number of networks of contacts and interactions. However, quantifying how reliable these data are remains an open problem. From Biology to Sociology and Economy, the identification of false and missing positives has become a problem that calls for a solution. In this work we extend one of newest, best performing models -due to Guimer\'a and Sales-Pardo in 2009- to directed networks. The new methodology is able to identify missing and spurious directed interactions, which renders it particularly useful to analyze data reliability in systems like trophic webs, gene regulatory networks, communication patterns and social systems. We also show, using real-world networks, how the method can be employed to help searching for new interactions in an efficient way.

\end{abstract}


\section{Introduction}

The last several years have witnessed many advances in what is today known as network science. Although the study of networks is not new, the availability of data in many different fields, ranging from techno-social systems to biological networks, has paved the way to solve relevant questions that were not accessible just a few years ago due to the lack of relevant data. It is however evident that we have advanced less in some fundamental questions, already put forward back in 2001 \cite{Strog2001}. One of such challenges is to understand how models, methods and results of networks theory change when one consider different kinds of links: directed or undirected, weighted or unweighted. Though there are many good examples of real networks than can be easily treated as undirected \cite{Barabasi}, probably there is an even larger number of systems in which links' directionality and/or weights make a difference. These systems include gene regulatory networks \cite{Bolouri,Babu1}, food webs \cite{foodwebs} or some interaction networks extracted from social media communication patterns \cite{politics,15M}.

On the other hand, the lack of data quality and complete information about interactions is an ubiquitous problem in most research areas where the framework of network modeling is applied. For example, classical social survey methods must deal with problems like sampling biases \cite{Kossinets}, or data loss \cite{Schafer,Butts}, which can compromise network-level analyses. The problem is even more acute when moving from social to biological systems like transcriptional regulatory maps, in which the promise of high-throughput biochemical techniques of revealing the system backbone (i.e., transcriptomes) has to deal with the inaccuracy that these methods often show. Microarray essays $-$the main tool to quantitatively measure the activity of large amounts of genes in a highly parallel fashion$-$ constitute a paradigmatic example of a powerful, but sometimes inaccurate or hardly reproducible technique \cite{Draghici,Technometrics,Ioannidis}. 

Focusing on the subfield of gene regulatory networks, one additional limitation to the network approach is the diversity -even conceptual- of the high number of different techniques used to infer regulatory interactions \cite{Plos,RegulonDB,BMC}. Lastly, the most important issue is probably the fact that the environmental conditions under which regulatory interactions take place are, in general, different for each interaction, and for a high proportion of cases only roughly known. This leads to the paradox that in many cases, reported regulations \cite{Plos,RegulonDB,subtilis} identified through very diverse experimental techniques, and under specific experimental conditions, are rarely similar when links identified through different experiments are compared. 

It is then of utmost importance to develop new ways to assess data reliability in complex directed networks. In this paper, we capitalize on a previous method proposed to study the very same problem but for undirected systems \cite{Guimera}. Specifically, we generalize the method proposed by Guimera \& Sales-Pardo \cite{Guimera} to the case in which links are directed, like in a regulatory network. By doing so, we are able to successfully identify missing and spurious interactions in several real-world networks. Finally, we test whether the method can be used to predict new links in a genome-wide transcriptional regulatory network \cite{Plos}, providing a robust methodology that could help and guide the experimental search for unnoticed regulations. 

\section{Results}
\subsection{The method}

Following \cite{Guimera}, let us suppose that we are working on a certain graph whose adjacency matrix is $A^o$, which is just an imperfect realization of a certainly ideal, ``true'' network $A$ to which we have no access. Being $X$ a certain measurable property of the network, we will call $p(X=x|A^o)$ the probability that, once observed the graph $A^o$, $X$ is equal to $x$ in the ideal system $A$. Then we have:
\begin{equation}
p(X=x|A^O)=\int_\mathit{M}p(X=x|m)p(m|A^O)dm
\label{px1}
\end{equation}
where $p(m|A^O)$ is the probability that $m$ is the model in a class $\textit{M}$ that gave the observation $A^O$, and $p(X=x|m)$ stands for the probability of model $m$ to generate networks in which $X=x$. As the term $p(m|A^O)$ is certainly difficult to estimate, we must reformulate the problem by using the Bayes theorem, to get:
\begin{equation} 
p(X=x|A^O)=\frac{\int_\mathit{M}p(X=x|m)p(A^O|m)p(m)dm}{\int_\mathit{M'}p(A^O|m')p(m')dm' }
\label{bayes}
\end{equation}
where $p(A^0|m)$ is the probability that $m$ gave $A^O$ among all possible adjacency matrices and $p(m)$ is the a priori probability of model $m$.

At this point, we need to select a class of models to integrate the former expression. The main hypothesis that lies beneath this method consists of assuming that the required family is that of stochastic-blocks-models (SBM). In the case of undirected networks, any of these SBM can be characterized by a partition $P$ of the set of nodes into blocks, and a probability matrix $\mathbf{Q}$ such that the element $Q_{\alpha,\beta}$ defines the probability that any of the nodes belonging to the block $\alpha$ be connected to any of the nodes within block $\beta$. So, the probability of two nodes being connected depends only on the blocks these nodes belong to within the partition $P$. Note that under these assumptions, $\mathbf{Q}$ is symmetric.

In order to deal with directed networks several possibilities are conceptually feasible. Here, we propose the following variation of the model. Instead of considering one single partition $P$ of the nodes' space, we will consider two partitions, $P_s$ and $P_r$. Every node $i$ must then belong, independently, to a block in each partition: $i\in\sigma_i$ with $\sigma_i\in P_s$ and $i\in\tau_i$ with $\tau_i\in P_r$. The partitions just take into account the fact that in directed networks, out-going and incoming links are treated separately. Thus out-going links of node $i$ will be determined by block $\sigma_i$ to which it belongs in the partition $P_s$. On its turn, and in an independent way, the in-degree will be given by the block $\tau_i$ in the other partition $P_r$ in which the node $i$ is located. Within this scheme, the probability of node $i$ sending a link to node $j$ is $Q_{\sigma_i,\tau_j}$. Remarkably, the probability of observing the opposite link is different, and equal to $Q_{\sigma_j,\tau_i}$.

This scheme, yet having the virtue of its computational tractability, conceptually captures the behavior of systems like transcriptional regulatory networks in which the statistics associated to in-degrees are very different to those regarding out-degrees \cite{Plos,RegulonDB}, being both relatively uncorrelated. This can be easily understood if one considers that the biochemical properties that define the susceptibility of a protein to be regulated by others are different to those that make the protein a regulator. While the information that will ultimately define the identity and the strength of the transcriptional regulations affecting a protein reside in its promoter region, its eventual ability to bind to the promoters of other target proteins depends on the presence and identity of a regulator domain within its protein sequence. Consequently, these two eventual roles of the protein are determined by DNA sequences that are independent and that, at least in principle, can evolve separately, both in prokaryotic \cite{Babu2} and eukaryotic cells \cite{Prudhomme}.

\subsection{Links reliabilities}

Each of the SBM is fully defined by determining the two partitions above and the probability matrix, hence $m=(P_s,P_r,\mathbf{Q})$. Additionally, we define the reliability of a certain link $i\rightarrow j$ as the probability:
\begin{equation}
R_{i,j}=P(A_{i,j}=1|A^o).
\end{equation}
On the other hand, the probability of observing the graph $A^o$ as a realization of a certain directed SBM is:
\begin{equation}
P(A^O|P_s,P_r,\mathbf{Q})=\prod_{\alpha\in P_s,\beta\in P_r}Q_{\alpha\beta}^{l^O_{\alpha\beta}}(1-Q_{\alpha\beta})^{r_{\alpha\beta}-l^O_{\alpha\beta}}
\end{equation}
Finally, the probability of observing a link in a network generated by one of these SBMs is given as:
\begin{equation}
P(A_{i,j}=1|P_s,P_r,\mathbf{Q})=Q_{\sigma_i,\tau_j}
\end{equation} 
with $\sigma_i\in P_s$ and $\tau_j\in P_r$. Substituting the three last expressions into Eq.\ \ref{bayes}, we get, after integration over all possible probability matrices for each case, the following expression for the reliabilities of links:
\begin{equation}
 R_{i\rightarrow j}=\frac{1}{Z}\sum_{\scriptsize{\begin{split}
{P_s\in\mathit{P_S}}\\
{P_r\in\mathit{P_R}}
\end{split}}} P(P_s,P_r)\frac{l^O_{\sigma_i,\tau_j}+1}{r_{\sigma_i,\tau_j}+2} e^{-H(P_s,P_r)}
\label{rel}
\end{equation}
with $P_S$ and $P_R$ standing, respectively, for the spaces of all possible partitions of nodes as link senders (S) and link receivers (R). In turn, $l^O_{\sigma_i,\tau_j}$ is the number of links observed between nodes placed in $\sigma_i$ in $P_s$, and nodes placed in $\tau_j$ in $P_r$. Regarding $r_{\sigma_i,\tau_j}$ it is the maximum possible value for $l^O_{\sigma_i,\tau_j}$, that is, the product of the sizes of blocks $\sigma_i\in P_s$ and $\tau_j\in P_r$. Finally, $P(P_s,P_r)$ is here the a priori probability of observing a subset of models defined by $P_s$ and $P_r$, under the assumption that once partitions are fixed, all possible models that one can get by changing the probability matrices are equally probable. In addition, the partition function $Z$ in the last equation takes the form:
\begin{equation}
 Z=\sum_{\scriptsize{\begin{split}
{P_s\in\mathit{P_S}}\\
{P_r\in\mathit{P_R}}
\end{split}}}P(P_s,P_r)e^{-H(P_s,P_r)}
\label{Z}
\end{equation}
and the hamiltonian function is:
\begin{equation}
H(P_s,P_r)=\sum_{\scriptsize{\begin{split}
{\alpha\in P_s}\\
{\beta\in P_r}
\end{split}}}\left[ ln\left(r_{\alpha\beta}+1)\right)+ln\binom{r_{\alpha\beta}}{l^O_{\alpha\beta}}\right ]
\label{H}
\end{equation}
Up to this point, the scheme of the method is totally analogous to the baseline method for undirected systems presented in \cite{Guimera}. However, the generalization of the method to directed networks requires further refinements. More precisely, as it is detailed in the Appendix, we must adopt here the following hypothesis. Let $\vec{\chi}_{P_x}$ be the vector whose components are the (ordered) number of nodes present in each of the blocks within partition $P_x$. We have that
\begin{eqnarray}
P(P_s,P_r)=constant\ \ \forall(P_s,P_r)\ \ with\ \  \vec{\chi}_{P_s}=\vec{\chi}_{P_r},\nonumber\\
P(P_s,P_r)=0\ \ \forall(P_s,P_r)\ \ with\ \  \vec{\chi}_{P_s}\neq\vec{\chi}_{P_r}.
\end{eqnarray}
Then, the a priori probabilities cancel out in Eqs.\ (\ref{rel}) and (\ref{Z}), and so, the mathematical forms of these expressions are identical to those given in \cite{Guimera}, except for the fact that here, sums and products are taken over the combination of two partition spaces: $P_s$ and $P_r$, with the additional constraint that the only couple of partitions $(P_s,P_r)$ that computes are those for which $\vec{\chi}_{P_s}=\vec{\chi}_{P_r}$ (See Appendix).

Nevertheless, the reliabilities sums have always the form of a canonical ensemble average, which allows us to use again a Metropolis algorithm to sample among all the possible pairs of partitions compatible with the condition $\vec{\chi}_{P_s}=\vec{\chi}_{P_r}$, those yielding to smaller hamiltonians and thus contributing the most to the sum (see Appendix). When the sampling finishes, we recover the reliabilities of all possible directed links in the network despite of their directionality --obviously, in general $R_{i,j}\neq R_{j,i}$--. Moreover, by ranking the links one can test which are the more reliable ones, no matter whether a given link was observed in our graph $A^{o}$ or not. This is what we do in the following sections.

\subsection{Method accuracy}

In order to check the performance of our approach, we perform a series of tests on top of different networks as in \cite{Guimera}. To this end, we use three well-known directed networks (see Appendix), namely: the trophic web of Narragansett bay, in the USA \cite{Narraganset}, the network compiled by Killworth and Bernard based on the radio calls recordings between a closed group of radio operators \cite{radio} and the directed synaptic wiring of the nematode \emph{C.elegans} \cite{elegans}.

\begin{figure*}[p]
\centering
  \includegraphics [angle=270,scale=0.8]{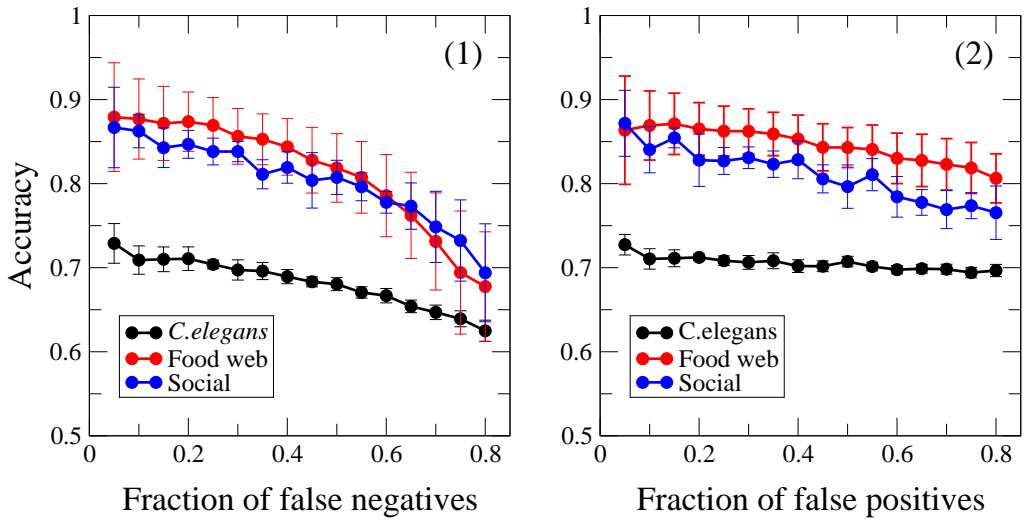}
\caption{Method accuracy in detection of missing (left) and bogus interactions (right) in three directed networks. The sampling processes are the standard described in the text for all the networks except the synaptic network of \emph{C.elegans}. In this case, (see Appendix) we take $T=10$ in the Metropolis algorithm, and we apply a threshold criterium in the sampling, considering only samples verifying $H>\langle H\rangle - \gamma \sigma_H$ with coefficient $\gamma\in [1,2]$, in order to reduce computational requirements.}
\label{fig1}
\end{figure*}

Assuming that these networks are error-free, we randomly remove a certain proportion of links. Then, we run our algorithm and rank the links reliabilities as coming out of the algorithm. We define the accuracy of the method when it comes to identify missing interactions as the probability that removed links are assigned a higher reliability ranking -i.e., they are false negatives- as compared to those that are true negatives. On the contrary, to test whether the method is able to identify bogus interactions accurately, we randomly add a proportion of links. As before, link reliabilities are computed and the ordered ranking is used to check the accuracy of the method, which in this case is given by the (mean) probability that bogus interactions -now they are like false positives- are ranked lower than true links. Results of the accuracy tests are shown in Fig.\ \ref{fig1}. As it can be seen, a good performance is obtained as compared with what one would expect by chance (i.e., accuracy equal to 0.5). Note, additionally, that the method performs qualitatively similar to the original algorithm developed for undirected systems \cite{Guimera}.  

\subsection{Guiding experiments}

Once we have tested the general performance of the model, we discuss its application in an important and specific domain, that of transcriptional regulatory networks. In this field of research, having a method like the one we are proposing here could help mitigate either the relatively poor quality and reduced size of some networks available \cite{Stathopoulos,Ojalvo} or to integrate vast amounts of information coming from high-throughput experimental techniques. 

On the other hand, there are several organisms, --even relevant pathogens-- for which the whole transcriptional map is not at hand, despite the fact that having the network would help in the search of new drug targets or vaccines. This is the case of the transcriptional regulatory network of \emph{Mycobacterium tuberculosis}. The bacillus of tuberculosis, responsible of one of the most threatening diseases worldwide, is probably one of the bacteria whose transcriptome has been best studied during the last years \cite{MtbRegList,Balazsi,Plos}. In 2008 the transcriptional regulatory network of the pathogen consisted of 782 genes and 937 interactions \cite{Balazsi}, but the last updated version, published in 2011, contains as many as 1624 genes and 3212 interactions \cite{Plos}. Moreover, the updated version, also added 357 new links between some of the 782 genes that were reported in 2008. 

All the aforementioned facts, together with the running costs of experiments are calling for methods that could optimize the search of new interactions. To test whether and to what extent our algorithm could contribute to cure new datasets and guide the experimental search of new transcriptional relations and regulators, we perform a simple exercise with the \emph{M. Tb} datasets of 2008 and 2011.

Specifically, we check whether the appearance of the 357 links in the 2011 compilation that connects pairs of genes already integrated in the 2008 network could have been inferred from the analysis of the 2008 network itself.
 
\begin{figure}[h]
\centering
  \includegraphics [angle=270,scale=0.8]{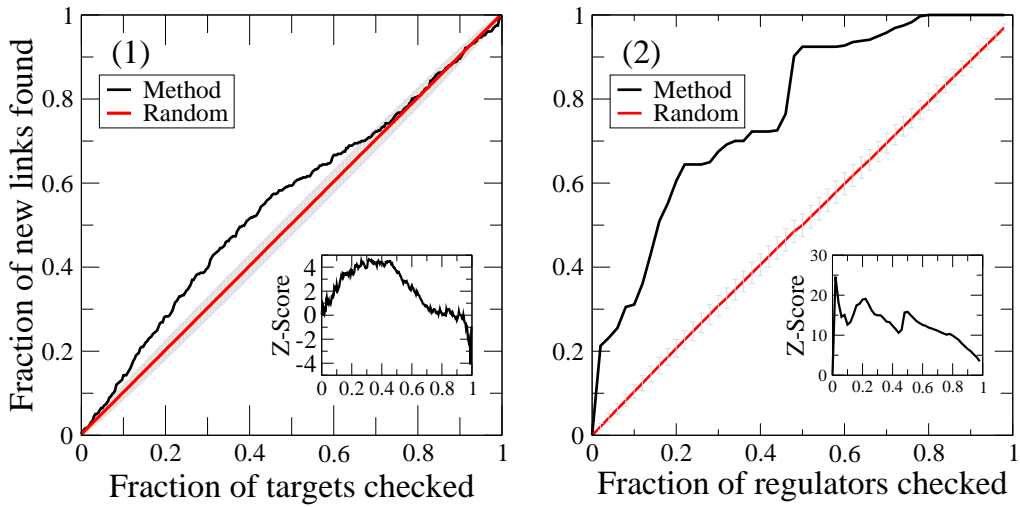}
\caption{\emph{Mycobacterium tuberculosis} transcriptional regulatory network update analysis. Panel 1: Regulators based search: Proportion of targets checked versus proportion of new links found, when focusing on regulators sending the new links. Panel 2: Number of regulators checked versus proportion of new links found, when focusing on targets receiving the new links. In the insets, the Z-Score of the methods' performance is computed, when compared to the random procedures, whose error bars ($\sigma=1$) are represented in grey. As it can be seen, the method outperforms the random procedure, mostly at first stages, and more remarkably in the case of target based search (panel 2)} 
\label{fig2}
\end{figure}

To simulate the way in which our method could help to identify these new interactions, let us suppose that we are interested on a certain gene of the 2008 network and we look for undiscovered regulations it might receive from any of the regulators already present in the network in 2008 --obviously excluding those that had been already found to regulate its activity at the moment--. If no biological clue is available about what regulators are the more likely candidates to act on our gene, we are forced to experimentally try, one after another, all the possibilities. If the result of some of these experiments is positive, and so the interactions exist, we will identify them at a linear rate, as it is represented in red in figure \ref{fig2}, panel 2. In the same figure, the black curve represents the rate at which all these novel interactions are detected when the possible targets are checked according to their reliabilities. As we can see, the proposed method greatly enhances the rate at which new links are discovered, which in practice could represent saving time and resources.

If the situation is the opposite, and we are interested on unveiling new regulations coming from any of the regulators of the network in 2008, the rate at which we will experimentally find the targets of the new links is represented in figure \ref{fig2}, panel 1, when choosing the candidates according to their reliabilities, and when the order is random. Again, we obtain that the performance of our method is consistently better than the random case, in this case only at the first stages of the search: starting from the regulators (Fig. \ \ref{fig2}, panel 2) and aiming at finding 50\% of the new targets, one has to seek the 37\% of the targets with the highest reliabilities in each case. This implies that the method proposed here uses 74\% of the time and resources needed if the identification is made randomly. Going back to the results shown in Fig. \ref{fig2}, panel 2, that case produces even better results: to find the 50\% of the regulations received by a target gene, one must only seek a 16\% of the total of regulators. Therefore, the method remarkably outperforms the random search by using as less as 32\% of the resources spent in the random case.

\section{Conclusions} 

We have proposed a generalization of the method in \cite{Guimera} to determine link reliabilities in directed networks. This opens the path to the potential application of our technique for the detection of missing and spurious interactions in systems as important as food-webs, transcriptional regulatory networks or certain social networks, all of which are directed networks. 

The accuracy and robustness of the method has been tested exhaustively on networks of different sizes and topological properties. Results of intensive numerical simulations have shown that missing and spurious interactions can be detected successfully. Additionally, we have numerically shown that the method can be used to guide the experimental search for missing links, as the reliability ranking resulting from the application of the algorithm to an incomplete network provides a very good guideline for experimental tests that eventually lead to the discovery of new interactions in a highly efficient way. 
This potentiality has important implications for our current efforts to map out transcriptional regulations, particularly, in cases such as that of {\em Mycobacterium tuberculosis}, where experimental lab protocols are very slow and expensive.  

Our model has however an important limitation. It is prohibitively costly in terms of computational time for large systems. Therefore, the method proposed here is mainly aimed at relatively small systems. For larger networks, the problem remains open and other solutions have to be found. Alternatively, we believe that the method presented in this paper could also be applied to subgraphs, overcoming in this way the size limitations. For instance, one can try to partition the whole system first by using one of the many algorithms available for community detection and then apply the reliability technique only to the detected communities. This kind of solutions will be explored in future work.

\section*{Appendix. Some relevant aspects regarding the methodology.}

\subsection{Phase space}

In \cite{Guimera}, the mathematical form of the hamiltonian, in the undirected model is, as said before, equivalent to \ref{H}, except for the fact that there is only a partition family to sum over.  Let us write it as:

\begin{equation}
H_u(P)=\sum_{\alpha<\beta}\left[ ln\left(r_{\alpha\beta}+1\right)+ln\binom{r_{\alpha\beta}}{l^O_{\alpha\beta}}\right ]
\label{Hps}
\end{equation} 

The restriction $\alpha<\beta$ (both blocks belonging to the partition $P$) appears only in order not to sum each term of the sum twice. Let's inspect the two different terms:
\begin{equation}
H_{1u}(P)=\sum_{\alpha<\beta} ln\left(r_{\alpha\beta}+1\right )
\label{H1}
\end{equation}  
\begin{equation}
H_{2u}(P)=\sum_{\alpha<\beta}ln\binom{r_{\alpha\beta}}{l^O_{\alpha\beta}} \label{H2}
\end{equation}  
The first term depends, essentially, on how ``concentrated'' the partition is. Briefly, it is minimal when the nodes tend to concentrate in a few number of blocks. In the case of having all the nodes on the same block, Eq.\ \ref{H1} gives $ln(1+N(N-1)/2)$, where $N$ is the number of nodes, which is approximately equal to $2ln(N)$ when $N$ is large enough. Instead, if we have the opposite situation in which each node is assigned to a different block, then $H_1=N(N-1)ln(2)>>2ln(N)$. So, the term $H_1$ minimizes when the partitions are compact, and maximizes in the opposite case. As for the second term, the picture is the opposite. The presence of the combinatory number implies that, to minimize $H_2$, the partitions of nodes should be a kind of ``straight fit'' for the links connecting blocks: given any two random blocks $\alpha$ and $\beta$, there should be a number of links between the blocks near to the maximum -the product of the block sizes, i.e. $r_{\alpha,\beta}$- or to the minimum (i.e. no link between the blocks). So, if we aim at getting the minimum of this term alone, one must go to the segregated partition in which each node belongs to a different block, for which the term directly vanishes. 

Therefore, minimizing the hamiltonian implies finding a compromise between aggregation and segregation of nodes into blocks, as the two terms have clear opposite effects, and no one of the extreme situations are globally convenient. How this picture change when we move to the bipartite scheme? The addition of new degrees of freedom to the system generates an undesirable situation in which, if we perform a Metropolis algorithm letting freely evolve the two partitions, we will reach a situation in which in the $P_s$ space, all nodes gather together into a single block, while in the $P_r$ space we will get an split into as many blocks as nodes are. The reason is that, for the system, such configuration is globally stable, because the two hamiltonian terms, under this configuration, reach values that are far away of the possible maximum. However, in this case, the final configuration is absolutely uninformative.

The above problem comes from the fact that the system is not constrained enough and it is allowed to adopt partitions in each one of the subspaces with very different degrees of aggregation. So, we should impose a further constraint so that the system can only adopt couples of partitions with the same aggregation state (i.e. $\vec{\chi}_{P_s}=\vec{\chi}_{P_r}$), the stable. This will allow to get partitions that give rise to minimum hamiltonians being at the same time fully informative and having a compromise at intermediate levels of aggregation between links assignments and block sizes. In this case, the algorithm will be qualitatively analogous to that of the undirected case. 

\subsection{Metropolis algorithm}

In order to perform our Metropolis algorithm, we start by assigning, at random, each node to one block, for example in the space $P_s$. Then we copy the partition generated to $P_r$. To ensure independence between the partitions but always verifying the constraint $\vec{\chi}_{P_s}=\vec{\chi}_{P_r}$, we proceed to randomize the partition $P_r$ by iteratively changing the block of couples of nodes (also chosen randomly) a high enough number of times. In this way, the blocks numbered equally in both partitions contains the same number of nodes. Thus, at each Metropolis step, we choose a couple of nodes belonging to the same block in both the partitions $P_s$ and $P_r$ and we try to change both at the same time to the same destination block (each one on its own partition). To ensure that any couples of nodes has the same probabilities of being chosen, we proceed as follows: we start by choosing randomly one node $n_1$ in one partition. Then we move it to the twin block containing the very same node $n_1$ in the complementary partition. Inside this twin block, we randomly choose the second node to move, $n_2$. After the nodes $n_1,n_2$ are selected and tentatively moved, we recalculate $H$ and accept the move if $H(t+1)<H(t)$. As usual, if the hamiltonian raises up, we accept the move with probability $P=e^{(H(t)-H(t+1))}$ in the standard case. Such an algorithmic scheme guarantees an ergodic exploration of the phase space, and ensures without problems detailed balance. In this way, after a certain transient, the hamiltonian reaches its equilibrium value and at that point, we start the sampling procedure, taking care that two consecutive samples are uncorrelated enough.

\subsection{Technical aspects}
 
While the method does not raise any problem when analyzing systems of small size (let us say $N<200$ nodes and $E<1000$ links approx.), as those studied in the preceding section, for larger networks, there sometimes appear some conceptual problems that can make the sampling procedure more difficult. First, it has been observed that the amplitude of oscillations of stationary hamiltonians, in general, increases with the size of the network analyzed. This range can be near $1000$ hamiltonian units for systems of less than $2000$ nodes, such as those of E.coli \cite{RegulonDB} or M.tuberculosis \cite{Plos} transcriptional regulatory networks. Since the distribution of the hamiltonians is qualitatively normal around the average value (results not shown), the higher the amplitude of the oscillation is, the lower the proportion of samples that will contribute significantly to the sum is (let us say, those with $H$, at most, 10 units greater than the minimum). This problem, when it comes to analyze big networks, will force us to get a too high number of samples to get a minimum amount of relevant ones. The latter can be prohibitive in terms of computational time (recall, in addition, that the computational time of a single Metropolis step also increases with the size of the system). 

\begin{figure}[h]
\centering
  \includegraphics[angle=270,scale=0.8]{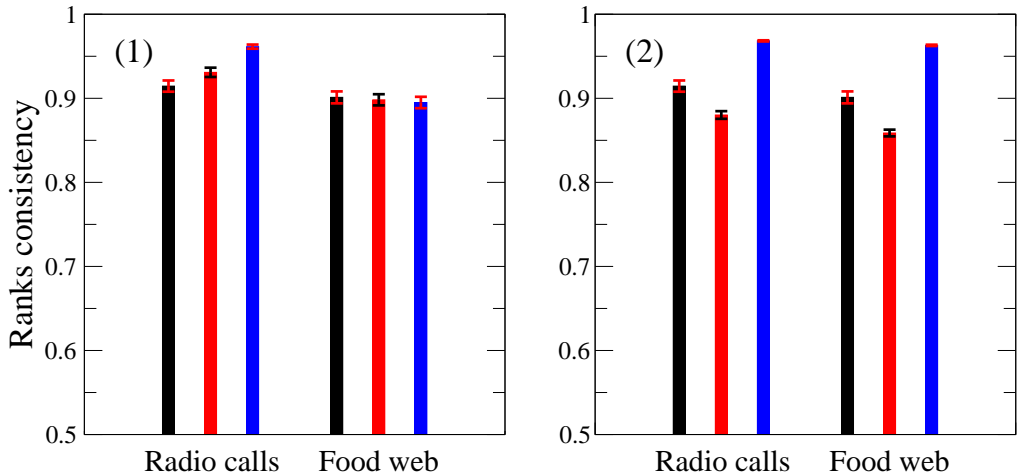}
\caption{Coherence of ranks defined as the proportion of reliabilities that preserve ordering in successive realizations obtained with diverse sampling strategies: Panel 1: black bars: standard sampling procedure. Blue bars: Threshold sampling. We have set $\gamma=2$. See the text for further details. Red bars: relative coherence of standard sampling ranks vs. threshold sampling ranks ($\gamma=2$). Panel 2: black bars: standard sampling procedure. Blue bars: Hot-threshold sampling.($T=2$, $\gamma=2$). Red bars: relative coherence of standard sampling ranks vs. Hot-threshold sampling ranks ($T=2$, $\gamma=2$). See the text and Appendix for details.}
\label{figS1}
\end{figure}

Here we propose an alternative procedure that can be implemented when the networks under study are too large and computational resources do not allow a full exploration of the phase space. The alternative is as simple as discarding all the samples with $H<\langle H_{stat}\rangle-\gamma\cdot\sigma_{H_{stat}}$, where $\gamma$ is a coefficient that can be chosen depending on the computational time we require and the number of samples we are looking for. This resource, although in principle could limit the performance of the method, does not affect it significantly, as showed in Fig.\ \ref{figS1}, panel 1. 

The black bars in figure \ref{figS1}, panel 1, show the consistency of the standard method of sampling without any threshold. We define this consistency as the proportion of reliabilities pairs $R_{i,j}$ $R_{k,l}$ whose relative ordering is preserved in successive reliability ranks obtained with the same method. Moreover, in red bars, the comparison is made between a rank obtained with the standard procedure and another rank for which only the samples that lie over a threshold $\langle H_{stat}\rangle-\gamma·\sigma_{H_{stat}}$ have been preserved and considered (here, $\gamma=2$). Finally, the bars in blue show the internal consistency of the threshold method, that is, the mean proportion of reliability pairs whose order is conserved when we compare pairs resulting from two independent rankings generated using the threshold criterium. As it can be seen, the three measures, for the two systems shown, are consistently high and quantitatively similar between them, thus providing evidence that the threshold method could help in situations where the required computational time is prohibitively large if we aim at getting enough samplings.

There is an additional problem that generally appears when the networks have high mean connectivities, or, strictly speaking, when the mean connectivity  is of the order of half the number of total possible links in the network, that is, in a directed network, $N^2/2$. In these cases, the information stored in the adjacency matrix is high, and so, being high the number of constraints, the dependency of the hamiltonian on the partitions defines a rough energy landscape that sometimes can become difficult to deal with. This situation can lead the system to fall into a local minimum after the thermalization process, and get trapped there. So, once arrived to the stationary state, if the basin of that local minimum is small, we will observe that the system is not able to uncorrelate sufficiently, and thus, even if its energy is small enough to consider it an acceptable minimum, our sampling will be very poor. One solution to this issue would be that of parallelizing the algorithm starting each parallel process from a random initial configuration. In this way, the process will ideally reach independent minima and thus the sampling would be $N$ times reacher, being $N$ the number of parallel processes.

In the above solution is is not possible, the strategy would be to introduce a pseudo temperature $T>1$ in the Metropolis algorithm, just to ensure the system is able to abandon local minima and explore the whole configurational space looking for other ones. The adoption of this strategy has the problem that, the higher is the temperature, the higher is also the oscillation of amplitudes of the stationary hamiltonian, and therefore the application of a threshold might also be needed.

In Fig.\ \ref{figS1}, panel 2, we show the consistency of our method when the above strategy is implemented (using $T=2$) in combination with a threshold criterium to select the samples, accepting only those with $H>\langle H_{stat}\rangle-2\sigma_{H_{stat}}$. Though these operations, again, could compromise the quality of our sampling, we found that the consistency of the ranks generated with the method (Fig.\ \ref{fig2}, panel 2, red bars) compared to those generated by the standard procedure is higher than 85\%. On its turn, when we check the internal consistency of the ranks generated with the method is even better and could be greater than that reached with a standard sampling. 

\subsection{Network models}
\begin{itemize}

\item {\em Narragansett bay trophic web}. The dataset \cite{Narraganset} contains originally 220 interactions between 35 nodes. We have removed the links involving the nodes associated to input, output and respiration fluxes, in order to take into consideration only the trophic relationships between organisms. The effective size of our system is, thus, 32 nodes and 158 links.

\item {\em Killworth-Bernard radio calls network}. In their work \cite{radio}, the authors asked to 44 radio operators (nodes) to rank from 0 to 9 the frequency they had used to call the rest of operators during last month. We have reconstructed our network by assigning a link when the rank associated to it was greater than 1, which produces 400 connections. 

\item {\em C. Elegans synaptic network}. Though the network can be analyzed as a non-directed graph by considering a link between neurons despite of the directionality of the interaction, synaptic interactions are asymmetric, thus the system should be analyzed as a directed network. The network has 279 nodes and 2990 links \cite{elegans}.

\end{itemize}

\section*{Acknowledgments}

J.S. was supported by MINECO through an FPU fellowship. E. C is funded by the FPI program of the Government of Arag\'on, Spain. This work has been partially supported by MINECO through Grant FIS2011-25167, Comunidad de Arag\'on (Spain) through a grant to the group FENOL and by the EC FET-Proactive Project MULTIPLEX (grant 317532).

\end{document}